\documentclass[a4paper,journal,10pt]{IEEEtran}
\usepackage{color}
\usepackage{cite}
\usepackage{amssymb}
\usepackage{amsmath}
\usepackage{graphicx}                  
\usepackage{bm}
\usepackage{epsfig,psfrag}
\usepackage{epstopdf}

\newcommand{\be}{\begin{equation}}
\newcommand{\ee}{\end{equation}}
\newcommand{\ist}{\hspace*{.3mm}}
\newcommand{\rmv}{\hspace*{-.3mm}}

\newcommand{\rrmv}{\hspace*{-1mm}}
\newcommand{\bd}[1]{\mathbf{#1}}
\newcommand{\cl}[1]{\mathcal{#1}}
\newcommand{\nn}{\nonumber}

\def\mathlette#1#2{{\mathchoice{\mbox{#1$\displaystyle #2$}}%
                               {\mbox{#1$\textstyle #2$}}%
                               {\mbox{#1$\scriptstyle #2$}}%
                               {\mbox{#1$\scriptscriptstyle #2$}}}}
\newcommand{\matr}[1]{\mathlette{\boldmath}{#1}}
\newcommand{\RR}{\mathbb{R}}

\allowdisplaybreaks

\begin{document}
\title{Cooperative Localization for Mobile Networks:\\
A Distributed Belief Propagation \!--\! Mean Field \\
Message Passing Algorithm\vspace{1mm} 
\thanks{Final manuscript, \today.}
\thanks{B.\ \c{C}akmak, T.\ Pedersen, and B.\ H.\ Fleury are with the Department of Electronic Systems, Aalborg University, Aalborg, Denmark 
(e-mail: \{buc, troels, bfl\}@es.aau.dk). D. N. Urup is with Danish Defence, Denmark (e-mail: dnu1984@gmail.com). F.\ Meyer is with
CMRE,
La Spezia, Italy (e-mail: florian.meyer@cmre.nato.int).
F.\ Hlawatsch is with the Institute of Telecommunications, 
TU Wien, Vienna, Austria (e-mail: franz.hlawatsch@nt.tuwien.ac.at). This work was supported by the European Commission in the framework 
of the FP7 Network of Excellence NEWCOM\# (grant 318306) and by the 
Austrian Science Fund (FWF) under grant P27370-N30.}}

\author{Burak \c{C}akmak, Daniel N. Urup, 
Florian Meyer, \IEEEmembership{Member,~IEEE}, Troels Pedersen, \IEEEmembership{Member,~IEEE},\\ 
Bernard H. Fleury, \IEEEmembership{Senior Member,~IEEE}, and Franz Hlawatsch,~\IEEEmembership{Fellow,~IEEE}\vspace*{-2mm}}

\maketitle

\begin{abstract}
We propose a hybrid message passing method for distributed cooperative localization and tracking of mobile agents. Belief propagation 
and mean field message passing are employed for, respectively, the motion-related and measurement-related part of the factor graph. 
Using a Gaussian belief approximation, only three real values per message passing iteration have to be broadcast to neighboring agents. 
Despite these very low communication requirements, the estimation accuracy can be comparable to that of particle-based belief propagation.
\end{abstract}

\begin{IEEEkeywords}
Belief propagation, mean field approximation, cooperative localization, distributed estimation, information projection, Kullback-Leibler-divergence, mobile agent network. 
\end{IEEEkeywords}

\vspace*{-2mm}
\section{Introduction}\label{sec:intro}

Cooperative localization is a powerful approach for mobile networks \cite{patwari, ihler, wymeersch,mazuelas09,dardari12}. 
An attractive methodology for cooperative localization is sequential Bayesian estimation via message passing algorithms 
\cite{loeliger}. In particular, distributed belief propagation (BP) message passing algorithms were proposed in 
\cite{ihler,wymeersch,caceres,lien,savic12reduc,meyer16coslat,meyer14sigma} to localize static or mobile agents. Feasible implementations 
involve certain approximations and use, e.g., particle methods \cite{ihler, wymeersch, lien, savic12reduc, meyer16coslat} 
or the sigma point technique \cite{meyer14sigma}. 
Each message transmitted between neigh\-boring agents is a set of hundreds 
or more particles in the former case \cite{ihler, wymeersch, lien} and a mean and a covariance matrix, i.e., five real numbers in 2-D localization, 
in the latter case. For static agents, also message passing algorithms based on expectation propagation \cite{welling07,samuel2012comparison} or
the mean field (MF) approximation \cite{pedersen} were proposed. 
Similarly to sigma point BP \cite{meyer14sigma}, they use
a Gaussian approximation and the transmitted messages consist of a mean and a covariance matrix. 

In this letter, building on the theoretical framework in \cite{riegler13}, we present a distributed hybrid BP--MF message passing method for 
cooperative localization and tracking of mobile agents. We employ BP and MF \cite{riegler13} for, respectively, the motion-related 
and measurement-related part of the underlying factor graph, and we use a Gaussian belief 
approximation. Each BP--MF iteration includes an information projection \cite{koller09} that is efficiently implemented by means of a Newton 
conjugate-gradient technique \cite{CG}. Our method can achieve an accuracy comparable to that of BP-based methods 
with the same communication cost as the MF method \cite{pedersen}, i.e., three real numbers per transmitted message in 2-D localization.

This letter is organized as follows. The system model is described in Section \ref{sec:sysModel}. The hybrid BP--MF scheme is developed in 
Section \ref{sec:messagePass}, and the Gaussian belief approximation in Section \ref{sec:gaussian}. Section \ref{sec:simRes} presents simulation results.

\section{System Model}
\label{sec:sysModel}

The mobile network at discrete time $n\!\!\in \!\!\{1,...,N\}$ is described by a set of network nodes $\mathcal V^n$ and a set of edges $\mathcal E^n$ 
representing the communication/measurement links between the nodes. The set $\mathcal V^n$ is partitioned into a set $\mathcal V_{{\rm M}}^n$ 
of mobile agents at unknown positions and a set $\mathcal V_{{\rm A}}^n$ of static anchors at known positions. An edge $(k,l)\in \mathcal E^n$ indicates 
the fact that agent or anchor $l$ transmits data to agent $k$ and, concurrently, agent $k$ acquires a noisy measurement of its distance to agent or anchor $l$. 
The edge set $\mathcal E^n$ is partitioned into a set $\mathcal E_{\rm M}^n$ of edges between certain agents, i.e.,  $(k,l) \!\in\! \mathcal E_{\rm M}^n$ 
implies $k,l \!\in\! \mathcal V_{{\rm M}}^n$, and a set $\mathcal E_{\rm MA}^n$ of edges between certain agents and anchors, i.e., 
$(k,l) \!\in\! \mathcal E_{\rm MA}^n$ implies $k \!\in\! \mathcal V_{{\rm M}}^n$ and $l \!\in\! \mathcal V_{{\rm A}}^n$. 
Information exchange between agents is bidirectional, i.e., $(k,l)\!\in\! \mathcal E_{\rm M}^n$ implies $(l,k)\in \mathcal E_{\rm M}^n$. 
We consider a distributed scenario where each agent knows only its own measurements. Since the anchors have exact knowledge of their own position, 
they do not need to acquire measurements and receive position information from neighboring nodes. Accordingly, 
anchors transmit position information to agents but not vice versa, i.e., $(k,l)\!\in\! \mathcal E_{\rm MA}^n$ implies $(l,k) \notin \mathcal E_{\rm MA}^n$.

Let the vector $\matr x_{k}^n$ denote the state  of agent $k\in \mathcal V_{\rm M}^n$ at time $n\!\!\in \!\!\{1,...,N\}$. 
Moreover, let $\matr x^{n}\rmv\triangleq {\big[\matr x_{k}^{n}\big]}_{k \in \mathcal{V}_{\rm M}^n}$ and $\matr x^{1:n}\triangleq\big[\matr x^{i}\big]_{i= 1}^n$. 
While our approach applies to any linear-Gaussian motion model, we here consider specifically those two motion models (MMs) that are most frequently 
used in practice. In MM1, $\matr x_{k}^n \rmv= \matr p_{k}^n \rmv\in \RR^2$ is the 2-D position of agent $k$ at time $n$. If agent $k$ belongs to the network 
at times $n$ and $n \rmv-\! 1$, i.e. $k\in\mathcal V_{{\rm M}}^n\cap\mathcal V_{{\rm M}}^{n-1}\rmv$, then $\matr p_{k}^n$ is assumed to evolve 
according to the Gaussian random walk model \cite{barShalom01} 
\[
\matr p_k^n=\matr p_k^{n-1}+\sqrt{T}\matr v_k^n \ist .
\]
Here, $T$ is the duration of one time step and $\matr v_k^n \rmv\in\rmv \RR^2$ is \nolinebreak 
zero-mean \nolinebreak 
Gaussian driving noise with component  variance $\sigma_{v}^{2}$. Note that $\matr v_k^n$ can be interpreted as a random velocity. 
In MM2, $\matr x_{k}^n =\big[(\matr p_{k}^{n})^\mathrm{T} \,\ist (\matr v_{k}^{n })^\mathrm{T} \big]^{\mathrm T}\!$, where $\matr v_{k}^n \rmv\in \RR^2$ 
is the 2-D velocity of agent $k$ at time $n$. For $k\in\mathcal V_{{\rm M}}^n\cap\mathcal V_{{\rm M}}^{n-1}\!$, $\matr x_{k}^n$ is assumed to evolve 
according to the constant velocity model \cite{barShalom01}
\begin{equation}
\matr x_k^n=\matr F\matr x_k^{n-1}+\matr G \matr a_k^n \ist.
\label{GRW-2}
\vspace{-1mm}
\end{equation}
Here, $\matr a_k^n \rmv\in \RR^2$ is zero-mean Gaussian driving noise (a random acceleration) with component variance $\sigma_{a}^{2}$. Moreover, 
$\matr F= {{\Big[\!\! {\small \begin{array}{cc}
	1 \!\!&\!\! T \\[-.1mm]
	0 \!\!&\!\! 1
	\end{array} } \rmv\!\rmv\Big] }\otimes \matr {\rm I}_2 }$
\vspace{-1.5mm}
and $\matr G= {\Big[ \!\rmv {\small \begin{array}{c}
T^2/2 \\[.2mm]
T
\end{array} } \!\!\Big] } \otimes \matr {\rm I}_2$,
where $\otimes$ denotes 
\vspace{1.5mm}
the Kronecker product and $\matr {\rm I}_m$ is the $m\! \times\! m$ identity matrix. 
Note that in both MM1 and MM2, the state-transition probability density function 
(pdf) $p(\matr x_k^{n}\vert \matr x_k^{n-1})$ is Gaussian. For agents that are part of the network at time $n$ but not at time $n-1$, i.e.,
$k \rmv\in\rmv \mathcal V_{{\rm M}}^{n} \rmv\setminus\rmv \mathcal V_{{\rm M}}^{n-1}\rmv$, we set $p(\matr x_k^{n}\vert \matr x_k^{n-1}) = p(\matr x_k^{n})$, 
where the prior pdf $p(\matr x_k^{n})$ is Gaussian. Under common statistical independence assumptions on 
$\matr v_k^n$ or $\matr a_k^n$ \cite{wymeersch}, the joint prior pdf of all agent states up to time $n$ is given by 
\vspace{-.1cm}
\begin{equation}
p(\matr x^{1:n}) \ist= \prod_{i=1}^{n} \prod_{k\in\mathcal V_{{\rm M}}^{i}} \! p \big(\matr x_k^{i}\vert \matr x_k^{i-1} \big) \ist. 
\label{prior}
\vspace{-1mm}
\end{equation}

If $(k,l)\in\mathcal E^n\!$, agent $k \rmv\in \cl{V}^{n}_{\rm M}$ acquires at time $n$ a noisy measurement of its distance to agent or anchor $l$,
\begin{equation}
d_{k,l}^n=\Vert\matr  p_{k}^n \!-\rmv \matr p_{l}^n \Vert + w_{k,l}^n \ist.
\label{meas_kodel}
\vspace{-.5mm}
\end{equation}
The measurement error $w_{k,l}^n$ is assumed zero-mean Gaussian with variance $\sigma_{w}^2$. 
Note that the local likelihood function $p(d_{k,l}^n|\matr{p}_{k}^n, \matr{p}_{l}^n )$ is nonlinear in $\matr{p}_{k}^n$ and $\matr{p}_{l}^n$.
Let $\matr d^{1:n} \!\triangleq\rmv \big[\matr d^{i}\big]_{i=1}^n\rmv$ with $\matr d^n \!\triangleq\! {\big[d_{k,l}^n\big]}_{(k,l)\in \mathcal E^n}$.
Assuming that all $w_{k,l}^n$ are independent, the global likelihood function involving all measurements and all states up to time $n$ factors according 
\vspace{-1mm}
to
\begin{align}
p(\matr d^{1:n}\vert \matr x^{1:n})=\prod_{i=1}^{n}\prod_{(k,l)\in \mathcal E_{\rm M}^{i}} \!\!\!\rmv p\big(d_{k,l}^{i}\vert \matr p_{k}^{i} , \matr p_{l}^{i}\big) 
\!\!\! \prod_{(\kappa,\lambda)\in \mathcal E_{\rm MA}^{i}} \!\!\!\!\!\rmv p\big(d_{\kappa,\lambda}^{i}\vert \matr p_{\kappa}^{i} , \tilde{\matr{p}}_{\lambda}^{i}\big), \nonumber\\[-3.5mm]
\label{likelihood}\\[-8mm]
\nonumber
\end{align}
where $\tilde{\matr{p}}_{\lambda}^n$ denotes the (known) position of anchor $\lambda \!\in\! \mathcal V_{\rm A}^n$.

\vspace{-.5mm}

\section{The Proposed Message Passing Scheme}
\label{sec:messagePass}

\vspace{.3mm}

The task of agent $k \rmv\in \cl{V}_{{\rm M}}^{n}$ is to estimate its state $\matr x^n_{k}$ from the total measurement vector $\matr d^{1:n}\rmv$,
for $n\in\{1,\ldots,N\}$. We will consider the minimum mean-square error (MMSE) estimator 
\begin{equation}
\hat{\matr x}^{n}_{k} \ist\triangleq \int \rmv\rmv \matr x^{n}_{k} \, p(\matr x_{k}^n\vert \matr d^{1:n}) \ist {\rm d}\matr x^{n}_{k} \, , \quad k \rmv\in \cl{V}_{{\rm M}}^{n} \ist .  
\label{eq:mmse}
\vspace{-1mm}
\end{equation}
Calculating the posterior pdf $p(\matr x_{k}^n\vert \matr d^{1:n})$
involved in \eqref{eq:mmse} by direct marginalization of the joint posterior pdf $p(\matr x^{1:n}\vert \matr d^{1:n})$ is infeasible because of the excessive 
dimension of integration and because $\matr d^{1:n}$ is not locally available at the agents. Next, we develop a distributed message passing scheme that 
approximates $p(\matr x_{k}^n\vert \matr d^{1:n})$, $k \rmv\in \cl{V}_{{\rm M}}^{n}$, $n\in\{1,\ldots,N\}$.

By Bayes' rule, $p(\matr x^{1:n}\vert \matr d^{1:n}) \propto p(\matr d^{1:n} \vert \matr x^{1:n}) \ist p(\matr x^{1:n})$, 
where $p(\matr x^{1:n})$ and  $p(\matr d^{1:n} \vert \matr x^{1:n})$ factor as in \eqref{prior} and \eqref{likelihood}, respectively.
This factorization underlies the proposed hybrid BP--MF 
message passing scheme, which provides approximate marginal posterior pdfs (``beliefs'') $q_k(\matr{x}_{k}^{n}) \rmv\approx\rmv p(\matr{x}_{k}^{n}|\matr d^{1:n})$ 
for all $k \!\in\! \cl{V}_{\rm M}^{n}$. Our scheme is an instance 
\pagebreak 
of the general hybrid BP--MF message passing scheme presented in \cite{riegler13}. 
We use BP for the motion-related factors $p(\matr x_k^{n}\vert \matr x_k^{n-1})$ and 
MF for the measurement-related factors 
$p(d_{k,l}^n\vert \matr p_{k}^n,\matr p_{l}^n)$, and we suppress all messages sent backward in time (cf.\ \cite{wymeersch}). 
We thus obtain the following iterative scheme at time $n$: In message passing iteration $t \in \{1,..., t^*\}$, beliefs $q_k^{[t]}(\matr x_{k}^n)$ are calculated as
\be
q_k^{[t]}(\matr x_{k}^n) \ist=\ist
\frac{1}{Z} \, m_{k\to k}(\matr x_{k}^n) \!\prod_{l\in \mathcal N_k^n} \!\! m^{[t]}_{l\to k}(\matr p_{k}^n) \ist , \quad k \rmv\in\rmv \cl{V}_{\rm M}^{n} \ist,
\label{1a}
\vspace{-1.5mm}
\ee
where $Z$ is a normalization constant and $\mathcal{N}_k^n \!\triangleq\rmv \{l \ist\vert\ist (k,l) \!\in\! \cl{E}^n\}$ is the set of agents and anchors communicating
with agent $k$ at time $n$ (termed ``neighbors''). The factors in \eqref{1a} are obtained 
\vspace{-3mm}
as 
\be
m_{k\to k}(\matr x_{k}^n) \ist= \begin{cases}\int \! q_k^{[t^*]}(\matr x_{k}^{n-1}) \ist p(\matr x_{k}^n\vert\matr x_{k}^{n-1}) \ist {\rm d}\matr x_{k}^{n-1}, \\[-.3mm]
\hspace{29mm} k\in\mathcal V_{{\rm M}}^n\cap\mathcal V_{{\rm M}}^{n-1}\\[.8mm]
p(\matr x_k^n) \ist,\quad k\in\mathcal V_{{\rm M}}^n \!\setminus\! \mathcal V_{{\rm M}}^{n-1}
\end{cases}
\label{2a} 
\vspace{-2mm}
\ee
and
\begin{align}
\hspace{-1.5mm}m^{[t]}_{l\to k}(\matr p_{k}^n) \ist=\ist \exp\!\bigg( \rmv\int \!\rmv q_l^{[t-1]}(\matr x_{l}^n) \ln p(d_{k,l}^n\vert \matr p_{k}^n,\matr p_{l}^n) 
  \ist {\rm d}\matr x_{l}^n \rmv \bigg) \ist . \!
\label{2b}
\end{align}
(Note that $\matr p_{l}^n \!=\! \tilde{\matr p}_{l}^n$ if $l$ is an anchor.) This recursion is initialized with $q_k^{[0]}(\matr x_{k}^n) = m_{k\to k}(\matr x_{k}^n)$. 

In a distributed implementation, each agent $k$ broadcasts its belief $q_k^{[t-1]}(\matr x_{k}^n)$ to its neighbors $l \in \mathcal{N}_k^n$ and receives the neighbor beliefs 
$q_l^{[t-1]}(\matr x_{l}^n)$, $l \in \mathcal{N}_k^n$. These beliefs are then used to calculate the messages $m^{[t]}_{l\to k}(\matr p_{k}^n)$,
$l \in \mathcal{N}_k^n$ at agent $k$ as in \eqref{2b}. These messages, in turn, are needed to calculate the updated belief 
$q_k^{[t]}(\matr x_{k}^n)$ at agent $k$ according to \eqref{1a}. After $t^*$ iterations, the final belief $q_k^{[t^*]}(\matr x_{k}^n)$ is used for state estimation, 
i.e. $q_k^{[t^*]}(\matr x_{k}^n)$ is substituted for $p(\matr x_{k}^n\vert \matr d^{1:n})$ 
\vspace{-0.1cm}
in \eqref{eq:mmse}.

\section{Gaussian Belief Approximation}\label{sec:gaussian}

Inspired by \cite[Section~IV]{pedersen}, we introduce an approximation of the message passing scheme \eqref{1a}--\eqref{2b} such that the beliefs 
are constrained to a certain class of Gaussian pdfs. This leads to a significant reduction of both interagent communication and computational complexity 
relative to a particle-based implementation. We first consider MM2. A more detailed derivation is provided in \cite{burak16supmat}.

\vspace{-0.2cm}

\subsection{Gaussian Belief Approximation for MM2}\label{sec:approx}

We constrain the beliefs to Gaussian pdfs by using the \emph{information projection} approach \cite{koller09}, i.e., substituting for $q_k^{[t]}(\cdot)$ in 
\vspace{-2mm}
\eqref{1a}
\begin{equation}
\tilde{q}_k^{[t]}(\cdot) \ist\triangleq\ist \operatorname*{\arg \min }_{g\in \mathcal G}\ist 
{D} \rmv \big[ g \ist\big\Vert q_k^{[t]} \big] \ist.
\label{1aa}
\end{equation}
Here, ${D}\big[g\Vert\ist q\big] \triangleq\int g(\matr x)\ln \frac{g(\matr x)}{q (\matr x)}\ist {\rm d}\matr x$ is the Kullback-Leibler divergence 
and $\mathcal G$ is the set of 4-D Gaussian pdfs $g(\matr x) ={N}(\matr x; \matr{\mu}, \matr C )$ with covariance matrix of the form
${\matr{C}}\rmv= \rmv {\Big[ \!\!\! \begin{array}{cc}
	c_{\text{p}} \!\!\!&\!\! c\\[-.3mm]
	c \!\!\!&\!\! c_{{\text{v}}}
	\end{array} \!\!\! \Big]} \otimes {\bf I}_2$.
We will denote the mean and covariance matrix of $\tilde{q}_k^{[t]}(\matr x_k^n) ={N}\big(\matr x_k^n; (\matr{\mu}_{k}^{n})^{[t]}, (\matr C_k^{n})^{[t]}\big)$ 
defined in \eqref{1aa} as
$(\matr{\mu}_{k}^{n})^{[t]}= {\left[\!\!\rmv{\small \begin{array}{c}
(\matr {\mu}_{\text{p},k}^{n})^{[t]} \\[.8mm]
(\matr {\mu}_{\text{v},k}^{n})^{[t]} 
\end{array}}\!\!\rmv\right] }$ and
$(\matr C_k^{n})^{[t]}= {\left[ \!\!\rmv {\small \begin{array}{cc}
	(c_{\text{p},k}^{n})^{[t]} \!\!\!&\!\rmv (c_{k~}^{n})^{[t]} \\[.8mm]
	(c_{k~}^{n})^{[t]}  \!\!\!&\!\rmv (c_{{\text{v},k}}^{n})^{[t]} 
	\end{array}} \!\!\rmv \right] } \rmv\otimes {\bf I}_2$.
Because direct computation of the minimizer \eqref{1aa} is infeasible, 
\pagebreak 
we resort to an iterative method. To that end, we first derive an analytical 
expression of the objective function ${D} \rmv \big[ g \ist\big\Vert q_k^{[t]} \big]$ in \eqref{1aa}, 
which we abbreviate by $F^{[t]}_k(\matr \theta)$ with $\matr\theta \triangleq [\matr\mu^\mathrm{T} \, c_{\text{p}} \; c_{\text{v}} \; c ]^\mathrm{T}\!$.
Using the factorization in \eqref{1a}, this function can be expressed as
\be
F_k^{[t]}(\matr\theta)  \,=\, D [ \ist g  \Vert\ist {m}_{k\to k}]
-\! \sum_{l\in \mathcal N_k^n} \!\rmv G_{k,l}^{[t]}(\matr{\mu}_{\text{p}}, {c}_{\text{p}}) + \gamma \ist,
\label{FGG}
\ee
where $\matr \mu_{\text{p}}$ is the 2-D vector consisting of the first two entries of $\matr \mu$, $\gamma$ is a constant, and
\begin{equation}
G^{[t]}_{k,l}(\matr{\mu}_{\text{p}}, {c}_{\text{p}}) \ist\triangleq\rmv \int\! \mathcal\cl{N}(\matr p_{k}^n; \matr{\mu}_{\text{p}}, {c}_{\text{p}}\bd{I}_2) 
\ist \ln {m}^{[t]}_{l\to k}(\matr p_{k}^n) \ist {\rm d}\matr p_{k}^n \ist.
\label{G_l}
\end{equation}
To derive an expression of $D [ \ist g  \Vert\ist {m}_{k\to k}]$
in \eqref{FGG}, we note that for $k\in\mathcal V_{{\rm M}}^n\cap\mathcal V_{{\rm M}}^{n-1}\!$, due to the Gaussian 
$\tilde{q}_k^{[t]}(\matr x_k^n)$ and the linear-Gaussian model \eqref{GRW-2}, the message in \eqref{2a} (in which $q_k^{[t^*]}(\matr x_k^{n-1})$ 
is replaced by $\tilde{q}_k^{[t^*]}(\matr x_k^{n-1})$) is also Gaussian, i.e., ${m}_{k\to k}(\matr x_{k}^n) = {N}( \matr{x}_{k}^n; \matr \eta^n_k, \matr \Sigma^n_k )$. 
By using \eqref{GRW-2} and standard Gaussian integral identities \cite{MCB}, we obtain for 
\vspace{.5mm}
$k\in\mathcal V_{{\rm M}}^n\cap\mathcal V_{{\rm M}}^{n-1}$
\begin{equation}
\hspace{.1mm}\matr \eta^n_k = \matr F \rmv(\matr{\mu}^{n-1}_k)^{[t^*]}\ist , \quad\! \matr \Sigma^n_k = \matr F (\matr{C}^{n-1}_{k})^{[t^*]} \matr F^{\mathrm{T}} 
\rmv+ \sigma_a^2 \matr{G} \matr{G}^{\mathrm{T}} \rmv.\hspace*{-1.5mm}
\label{2d}
\vspace{.3mm}
\end{equation}
For $k\in\mathcal V_{{\rm M}}^n \!\setminus\! \mathcal V_{{\rm M}}^{n-1}\!$, $\matr \eta^n_k$ and $\matr \Sigma^n_k$ equal, respectively, the mean and 
covariance matrix of the Gaussian prior $p(\matr x_k^n) = {N}(\matr x_k^n; \matr \eta^n_k, \matr \Sigma^n_k )$. Accordingly, we obtain in either case \cite{MCB}
\begin{align}
\hspace{-2mm}D [ \ist g  \Vert\ist {m}_{k\to k}]&=\ist \frac{1}{2} \ist \big[{\rm tr}\big((\matr \Sigma^n_k)^{-1}\matr{C}\big)
-\ln\det(\matr{C})\nn\\[-1mm]
& \qquad\;\; + (\matr{\mu} \!-\! \matr \eta^n_k )^\mathrm{T} (\matr \Sigma^n_k)^{-1} (\matr{\mu}\!-\! \matr \eta^n_k ) \big] + \gamma' \rmv ,
\label{KL2}
\end{align}
where $\gamma'$ is a constant. Furthermore, one can express $G^{[t]}_{k,l}(\matr{\mu}_{\text{p}}, {c}_{\text{p}})$ in \eqref{G_l} 
via an expectation of $-(d_{k,l}^n- \Vert \matr z_{k,l}^n\Vert)^2/\sigma_w^2$, where $\matr z_{k,l}^n$ is a 2-D Gaussian random vector with mean 
$\matr \mu_{\text{p}}-(\matr{\mu}^{n}_{\text{p},l})^{[t-1]}$ and variance $c_{\text{p}}+(c^{n}_{\text{p},l})^{[t-1]}$.
For $l\in {\mathcal V}^n_{\rm A}$, in particular, $(\matr{\mu}^{n}_{\text{p},l})^{[t-1]} = \tilde{\matr{p}}_{l}^n$ and $(c^{n}_{\text{p},l})^{[t-1]}=0$. 
By using expressions of the first-order and second-order moments of the Rician pdf \cite{Rice}, one obtains \cite{burak16supmat}
\begin{align}
&\hspace*{-3mm}G^{[t]}_{k,l}(\matr{\mu}_{\text{p}}, {c}_{\text{p}}) \nn \\[.7mm] 
&\hspace*{-2mm}=\ist -\frac{d_\mu^2 \rmv+ 2{c}_{\text{p}}}{2\sigma_w^2} +\frac{d^n_{k,l}}{\sigma^2_{w}}\sqrt{ \frac{\pi C}{2} \ist }  
  \ist M\bigg( \rmv {-\frac{1}{2}}\ist ;1\ist ; -\frac{d_\mu^2}{2 \ist C}\bigg) + \gamma''  , \!\!\!\label{f2} 
\\[-5mm]
\nonumber
\end{align}
where $d_\mu\rmv \triangleq \big\Vert \matr{\mu}_{\text{p}} \rmv-(\matr{\mu}^{n}_{\text{p},l})^{[t-1]} \big\Vert$, 
$C \triangleq c_{\text{p}} + (c^{n}_{\text{p},l})^{[t-1]}$, $M( \ist \cdot \, ; \ist\cdot \, ; \cdot \ist)$ 
denotes the confluent hypergeometric function of the first kind \cite{milton64}, and $\gamma''$ is a constant.

\subsection{Iterative Minimization
Algorithm for MM2} \label{sec:iterative}

\vspace{.3mm}

To derive an iterative algorithm for computing an approximation of 
$(\matr\theta_k^{n})^{[t]} \!=\! \big[(\matr {\mu}_{k}^{n})^{[t]\mathrm{T}} \, (c_{\text{p},k}^{n})^{[t]} \,\ist (c_{\text{v},k}^{n})^{[t]} \,\ist (c_{k}^{n})^{[t]} \big]^\mathrm{T}\!$,
i.e., of the minimizer of \eqref{FGG}, we set the gradient of $F_k^{[t]}(\matr \theta)$ to zero. This yields the following system of non-linear fixed-point equations 
$\matr \theta= (\chi_k^{n})^{[t]}(\matr \theta)$, whereof $(\matr \theta_k^{n})^{[t]}$ is a solution:
\begin{align}
\matr{\mu} &\ist=\ist \matr{\eta}_{k}^{n} + \matr \Sigma_k^{n} \!\sum_{l\in \mathcal N^n_k}  \!\rmv \frac{\partial G_{k,l}^{[t]}(\matr{\mu}_{\text{p}}, {c}_{\text{p}})}{\partial{\matr{\mu}}} \ist, 
  \label{G1}\\[-.5mm]
c_{\text{p}} &\ist=\ist \frac{c^2}{c_{\text{v}}} + \bigg(\frac{J_{k,11}^{n} \rmv+ J_{k,22}^{n}}{2} 
  \ist-\! \sum_{l\in \mathcal N_{k}^n} \!\rmv \frac{\partial G_{k,l}^{[t]}(\matr{\mu}_{\text{p}}, {c}_{\text{p}})}{\partial c_{\text{p}}}\bigg)^{\!\!-1} \rrmv, \label{G2} \\[.3mm]
c_{\text{v}}&\ist=\ist \frac{ c^2}{c_{\text{p}}} + \frac{2}{J^{n}_{k,33} \rmv+ J^{n}_{k,44}}  \ist ,
\label{G3}\\[1.5mm]
c &\ist=\ist \frac{1+\sqrt{1+ (J^{n}_{k,13}+J^{n}_{k,24})^2 \ist c_{\text{p}} \ist c_{\text{v}}}} {J^{n}_{k,13}+J^{n}_{k,24}} \ist ,
\label{G4} 
\end{align}
with $J^{n}_{k,ij} \triangleq \big[ (\matr \Sigma_k^{n})^{-1} \big]_{ij}$. The partial derivatives in \eqref{G1} and \eqref{G2} can be calculated 
using the relation  $\frac{{\rm d}M(-1/2\ist;1;\ist x) }{{\rm d}x} = -M(1/2 \ist ;2\ist ;x)/2$ \cite{milton64}, where $M(-1/2\ist;1;\ist x)$ can be computed 
efficiently via an approximation \cite[Section 4.5]{Daniel}.

A Newton conjugate-gradient method \cite[Chapter~7.1]{CG} is now applied to \eqref{G1}--\eqref{G4} to 
solve the system $\matr \theta= (\chi_k^{n})^{[t]}(\matr \theta)$ in $j_\text{max}$ steps, starting from an initial value $\matr \theta_0$. 
The method iteratively computes $\matr \theta_{j+1}=(\matr {\bf I}_{7}-\matr \Psi_j)\ist\matr\theta_{j} +\matr \Psi_j(\chi_k^{n})^{[t]}(\matr \theta_{j})$, 
where $\matr \Psi_j$ is the inverse of the Hessian matrix of $F_k^{[t]}(\matr\theta)$ at $\matr\theta_{j}$.
The Hessian matrix is approximated via the conjugate gradient, which requires only $F_k^{[t]}(\matr\theta)$
and its gradient \cite{CG}. While the algorithm's convergence has not been proven so far, it is suggested by our simulations. The 
algorithm may produce a local minimum of $F_{k}^{[t]}(\matr \theta)$, since this function is not convex 
in general. Therefore, the algorithm is run several times with different values of $\matr \theta_0$, and the result yielding
the smallest value of $F_{k}^{[t]}(\matr \theta)$ is retained. In our simulations, we used the generic routine \verb+scipy.optimize.fmin_tnc+ \cite{web}. 

\vspace{-2mm}

\subsection{Gaussian Belief Approximation for MM1}\label{sec:randomWalk}

The results in Sections \ref{sec:approx} and \ref{sec:iterative} can be used with minor changes also for MM1. We here have 
$\matr \mu \rmv=\matr {\mu}_{\text{p}}$ and $\matr C\rmv=$\linebreak 
${c}_{\text{p}} \bd{I}_2$, and the Gaussian belief approximation reads 
$\tilde{q}_k^{[t]}(\matr p_{k}^{n})={N}\big(\matr p_{k}^n; (\matr {\mu}_{\text{p},k}^{n})^{[t]}, ({c}^{n}_{\text{p},k})^{[t]} \bd{I}_2\big)$. The objective function 
$F^{[t]}_k(\matr \theta)$ (with $\matr\theta \triangleq [\matr\mu_{\text{p}}^\mathrm{T} \, c_{\text{p}} ]^\mathrm{T}$) is still given by \eqref{FGG} 
together with \eqref{KL2} and \eqref{f2}; however, the expressions \eqref{2d} are replaced 
\vspace{-.5mm}
by
\be
\matr \eta^n_k = (\matr {\mu}_{k}^{n-1})^{[t^*]} \ist,\quad 
\matr \Sigma^n_k = (\matr C_k^{n-1})^{[t^*]} + T \sigma_v^2 \ist{\bf I}_2 \ist, 
\label{2d_MM1} 
\vspace{-.8mm}
\ee
where $(\matr {\mu}_{k}^{n-1})^{[t^*]} \rmv=\rmv (\matr {\mu}_{\text{p},k}^{n-1})^{[t^*]}$ and $(\matr C_k^{n-1})^{[t^*]} \rmv=\rmv ({c}^{n-1}_{\text{p},k})^{[t^*]} \ist \bd{I}_2$.
Finally, fixed point equations in $\matr{\mu}_{\text{p}}$ and ${c}_{\text{p}}$ are obtained by setting to zero the gradient of $F_{k}^{[t]}(\matr \theta)$, 
and an iterative belief approximation algorithm is again based on these equations.

\vspace{-2.5mm}

\begin{figure*}[t]
\vspace{-.5mm}
\centering
\begin{minipage}[H!]{0.31\textwidth}
\psfrag{s01}[l][l][0.9]{\raisebox{-89mm}{\hspace{25.3mm}{(a)}}}
\psfrag{s02}[t][t][0.7]{\color[rgb]{0,0,0}\setlength{\tabcolsep}{0pt}\begin{tabular}{c}\raisebox{-1.5mm}{$\tau$}\end{tabular}}
\psfrag{s03}[b][b][0.7]{\color[rgb]{0,0,0}\setlength{\tabcolsep}{0pt}\begin{tabular}{c}\vspace{2mm}{$\hat{P}_{\text{out}}$}\end{tabular}}
\psfrag{s06}[l][l][0.55]{\color[rgb]{0,0,0}}
\psfrag{s07}[l][l][0.6]{\color[rgb]{0,0,0}SBP ($t^* \!=\rmv 30$)}
\psfrag{s08}[l][l][0.6]{\color[rgb]{0,0,0}SBP ($t^* \!=\rmv 5$)}
\psfrag{s09}[l][l][0.6]{\color[rgb]{0,0,0}BPMF ($t^* \!=\rmv 5$)}
\psfrag{s10}[l][l][0.6]{\color[rgb]{0,0,0}NBP ($t^* \!=\rmv 5$)}
\psfrag{s11}[l][l][0.6]{\color[rgb]{0,0,0}NBP ($t^* \!=\rmv 30$)}
\psfrag{s12}[l][l][0.6]{\color[rgb]{0,0,0}BPMF ($t^* \!=\rmv 30$)}
\psfrag{s14}[][]{\color[rgb]{0,0,0}\setlength{\tabcolsep}{0pt}\begin{tabular}{c} \end{tabular}}
\psfrag{s15}[][]{\color[rgb]{0,0,0}\setlength{\tabcolsep}{0pt}\begin{tabular}{c} \end{tabular}}

\psfrag{x01}[t][t][0.65]{$0$}
\psfrag{x02}[t][t][0.65]{$0.5$}
\psfrag{x03}[t][t][0.65]{$1$}
\psfrag{x04}[t][t][0.65]{$1.5$}
\psfrag{x05}[t][t][0.65]{$2$}
\psfrag{x06}[t][t][0.65]{$2.5$}
\psfrag{x07}[t][t][0.65]{$3$}
\psfrag{x08}[t][t][0.65]{$3.5$}
\psfrag{x09}[t][t][0.65]{$4$}

\psfrag{v01}[r][r][0.65]{$10^{-1}$}
\psfrag{v02}[l][l][0.65]{\raisebox{0mm}{\hspace*{-2.7mm}{$10^{0}$}}}

\centering
\hspace*{2mm}\includegraphics[height=3.5cm,width=\textwidth]{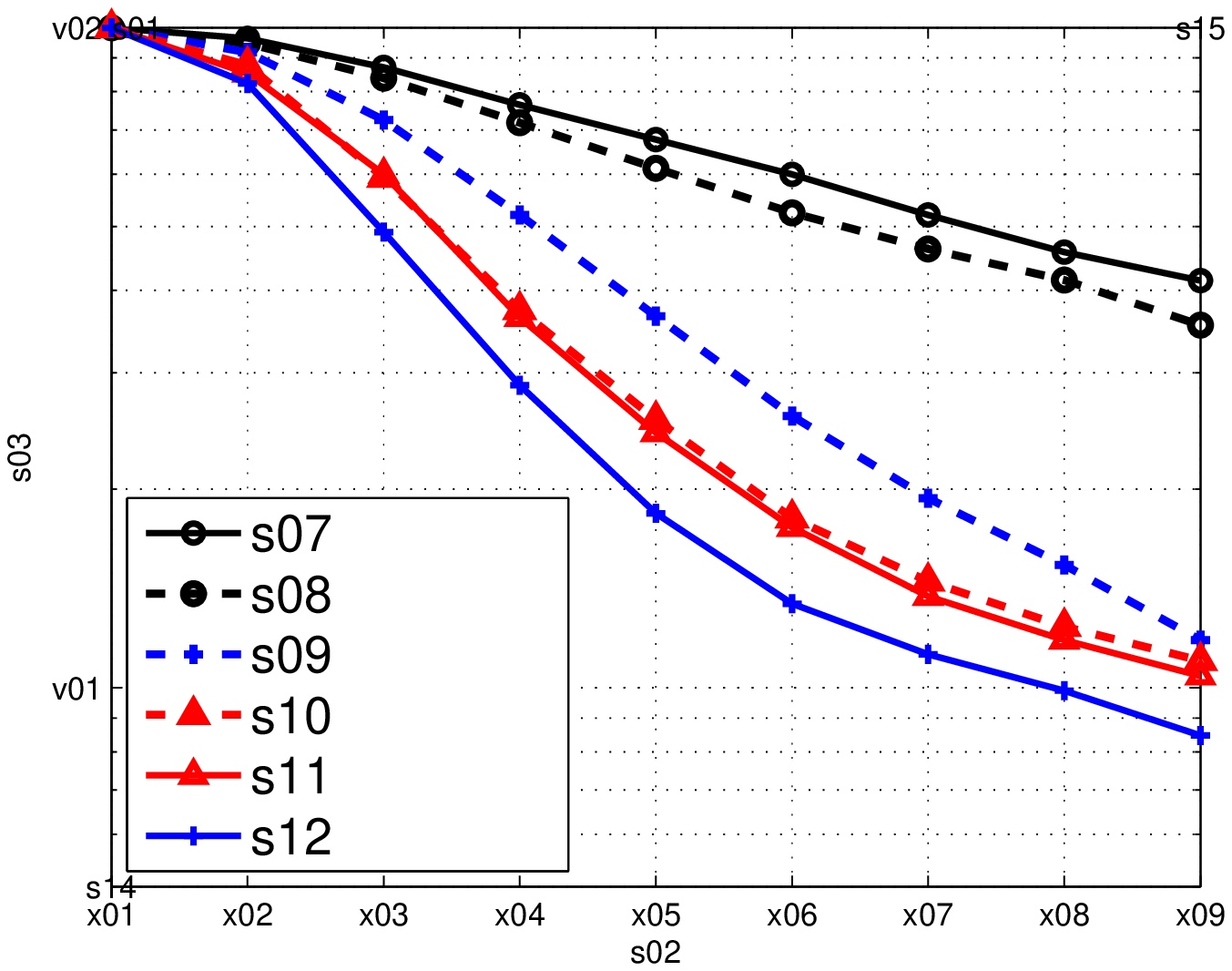}
\end{minipage}\hspace{3mm}
\begin{minipage}[H!]{0.31\textwidth}
\psfrag{s01}[l][l][0.9]{\raisebox{-89mm}{\hspace{25.3mm}{(b)}}}
\psfrag{s02}[t][t][0.7]{\color[rgb]{0,0,0}\setlength{\tabcolsep}{0pt}\begin{tabular}{c}\raisebox{-1.5mm}{$\tau$}\end{tabular}}
\psfrag{s03}[b][b][0.7]{\color[rgb]{0,0,0}\setlength{\tabcolsep}{0pt}\begin{tabular}{c}\vspace{2.7mm}{$\hat{P}_{\text{out}}$}\end{tabular}}
\psfrag{s06}[l][l][0.55]{\color[rgb]{0,0,0}}
\psfrag{s07}[l][l][0.6]{\color[rgb]{0,0,0}NBP ($t^* \!=\rmv 30$)}
\psfrag{s08}[l][l][0.6]{\color[rgb]{0,0,0}BPMF ($t^* \!=\rmv 5$)}
\psfrag{s09}[l][l][0.6]{\color[rgb]{0,0,0}BPMF ($t^* \!=\rmv 30$)}
\psfrag{s10}[l][l][0.6]{\color[rgb]{0,0,0}SBP ($t^* \!=\rmv 30$)}
\psfrag{s11}[l][l][0.6]{\color[rgb]{0,0,0}SBP ($t^* \!=\rmv 5$) }
\psfrag{s12}[l][l][0.6]{\color[rgb]{0,0,0}NBP ($t^* \!=\rmv 5$)}
\psfrag{s14}[][]{\color[rgb]{0,0,0}\setlength{\tabcolsep}{0pt}\begin{tabular}{c} \end{tabular}}
\psfrag{s15}[][]{\color[rgb]{0,0,0}\setlength{\tabcolsep}{0pt}\begin{tabular}{c} \end{tabular}}

\psfrag{x01}[t][t][0.65]{$0$}
\psfrag{x02}[t][t][0.65]{$0.5$}
\psfrag{x03}[t][t][0.65]{$1$}
\psfrag{x04}[t][t][0.65]{$1.5$}
\psfrag{x05}[t][t][0.65]{$2$}
\psfrag{x06}[t][t][0.65]{$2.5$}
\psfrag{x07}[t][t][0.65]{$3$}
\psfrag{x08}[t][t][0.65]{$3.5$}
\psfrag{x09}[t][t][0.65]{$4$}

\psfrag{v01}[r][r][0.65]{$10^{-2}$}
\psfrag{v02}[r][r][0.65]{$10^{-1}$}
\psfrag{v03}[l][l][0.65]{\raisebox{0mm}{\hspace*{-2.7mm}{$10^{0}$}}}

\centering
\hspace*{3mm}\includegraphics[height=3.5cm,width=\textwidth]{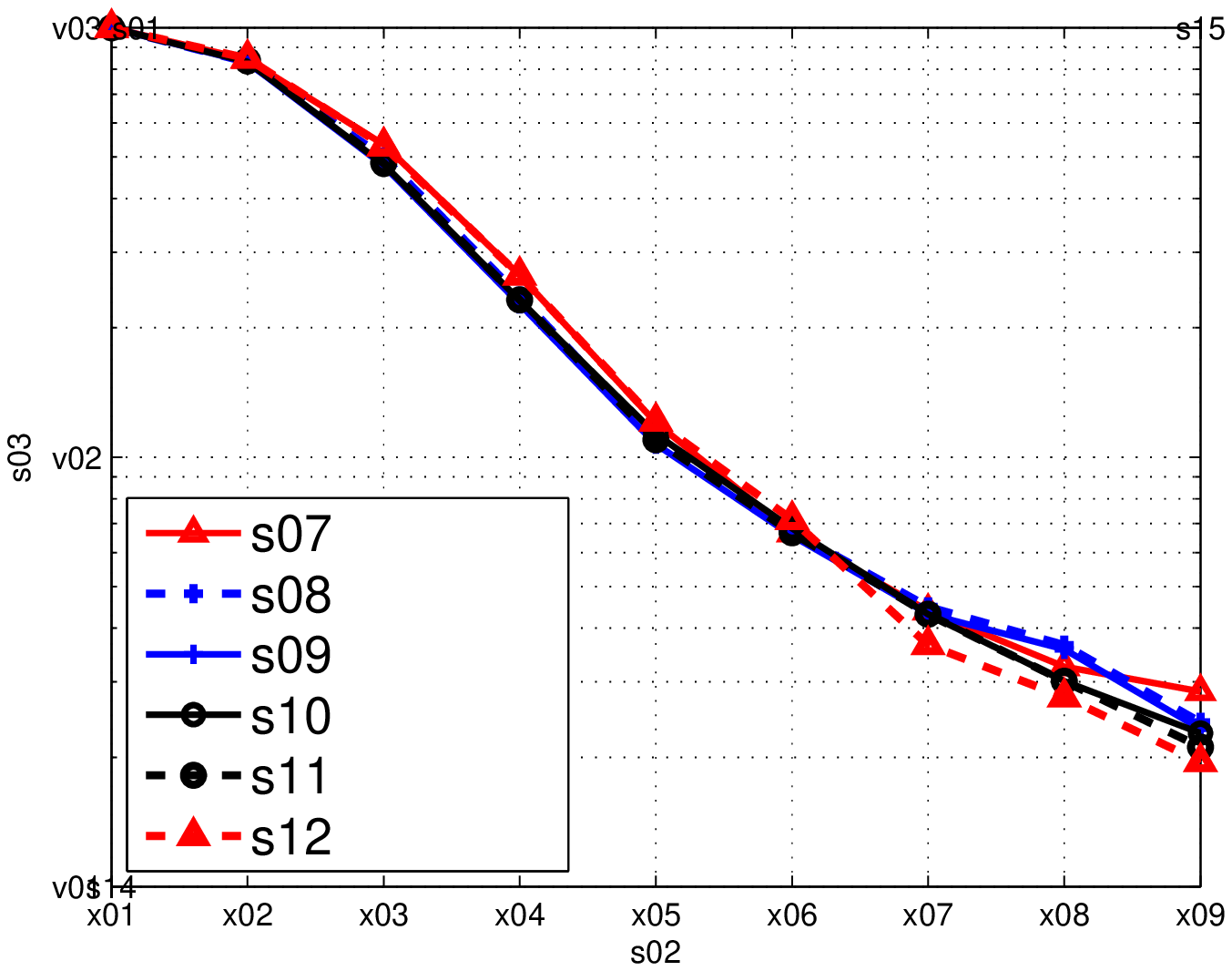}
\end{minipage}\hspace{3mm}
\begin{minipage}[H!]{0.31\textwidth}
\psfrag{s01}[l][l][0.9]{\raisebox{-89mm}{\hspace{25.3mm}{(c)}}}
\psfrag{s02}[t][t][0.7]{\color[rgb]{0,0,0}\setlength{\tabcolsep}{0pt}\begin{tabular}{c}\raisebox{-1.5mm}{$\tau$}\end{tabular}}
\psfrag{s03}[b][b][0.7]{\color[rgb]{0,0,0}\setlength{\tabcolsep}{0pt}\begin{tabular}{c}\vspace{2.7mm}{$\hat{P}_{\text{out}}$}\end{tabular}}
\psfrag{s06}[l][l][0.55]{\color[rgb]{0,0,0}}
\psfrag{s07}[l][l][0.6]{\color[rgb]{0,0,0}NBP ($t^* \!=\rmv 30$)}
\psfrag{s08}[l][l][0.6]{\color[rgb]{0,0,0}NBP ($t^* \!=\rmv 5$)}
\psfrag{s09}[l][l][0.6]{\color[rgb]{0,0,0}BPMF ($t^* \!=\rmv 5$)}
\psfrag{s10}[l][l][0.6]{\color[rgb]{0,0,0}BPMF ($t^* \!=\rmv 30$)}
\psfrag{s11}[l][l][0.6]{\color[rgb]{0,0,0}SBP ($t^* \!=\rmv 5$)}
\psfrag{s12}[l][l][0.6]{\color[rgb]{0,0,0}SBP ($t^* \!=\rmv 30$)}
\psfrag{s14}[][]{\color[rgb]{0,0,0}\setlength{\tabcolsep}{0pt}\begin{tabular}{c} \end{tabular}}
\psfrag{s15}[][]{\color[rgb]{0,0,0}\setlength{\tabcolsep}{0pt}\begin{tabular}{c} \end{tabular}}

\psfrag{x01}[t][t][0.65]{$0$}
\psfrag{x02}[t][t][0.65]{$0.5$}
\psfrag{x03}[t][t][0.65]{$1$}
\psfrag{x04}[t][t][0.65]{$1.5$}
\psfrag{x05}[t][t][0.65]{$2$}
\psfrag{x06}[t][t][0.65]{$2.5$}
\psfrag{x07}[t][t][0.65]{$3$}
\psfrag{x08}[t][t][0.65]{$3.5$}
\psfrag{x09}[t][t][0.65]{$4$}

\psfrag{v01}[r][r][0.65]{$10^{-2}$}
\psfrag{v02}[r][r][0.65]{$10^{-1}$}
\psfrag{v03}[l][l][0.65]{\raisebox{0mm}{\hspace*{-2.7mm}{$10^{0}$}}}

\centering
\hspace*{4mm}\includegraphics[height=3.5cm,width=\textwidth]{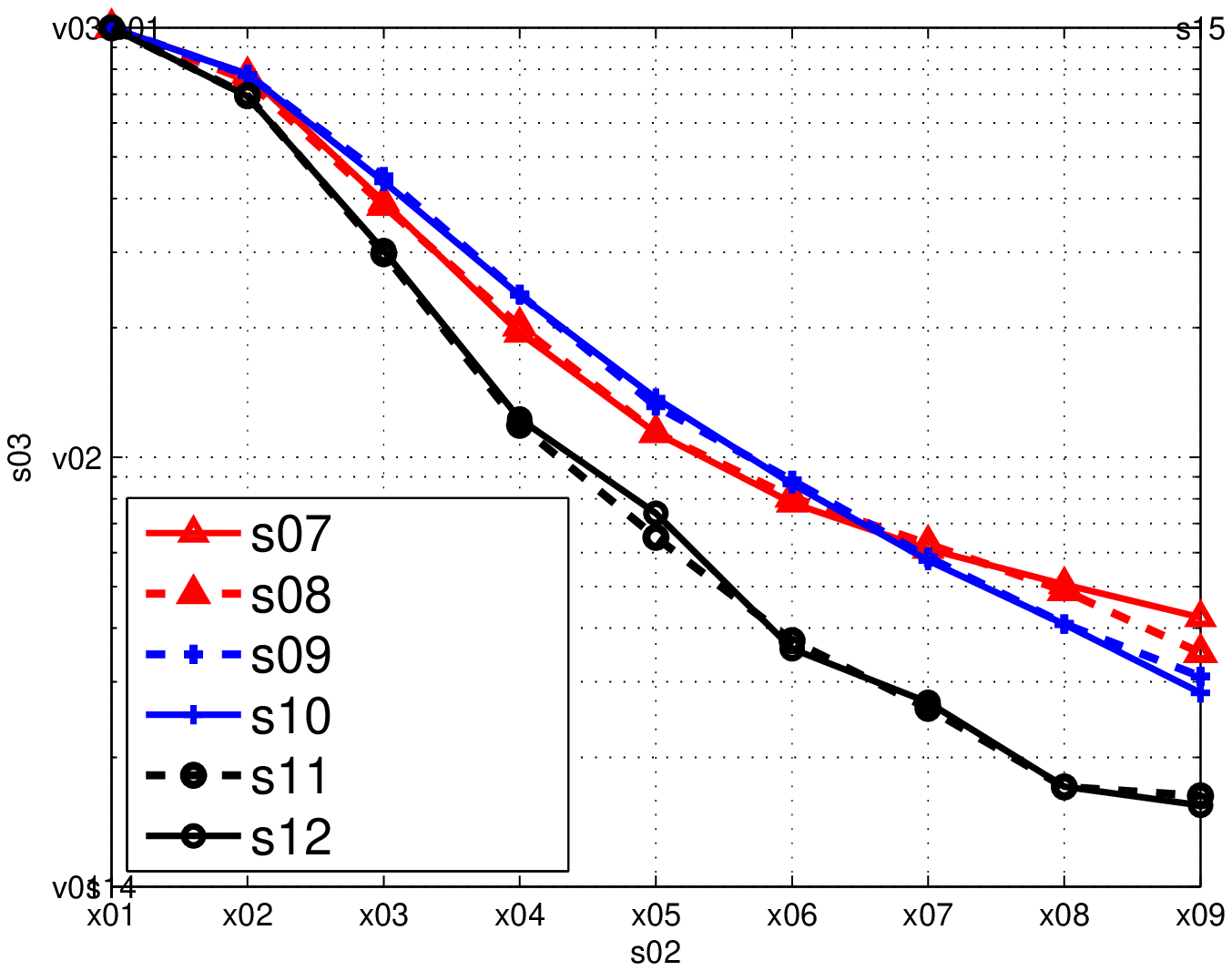}
\end{minipage}
\vspace{7mm}
\caption{Average outage probability versus outage threshold: (a) at $n \rmv=\! 1$ for both MMs, (b) at $n \rmv=\rmv 30$ for MM1, and (c) at $n \rmv=\rmv 30$ for MM2.} 
\label{fig:results}
\vspace{-.5mm}
\end{figure*}

\subsection{Distributed Cooperative Localization Algorithm} \label{sec:algorithmStatement}

The results of the previous subsections lead to a distributed algorithm for cooperative localization in which only parameters of Gaussian pdfs have to be 
communicated. At time $n$, agent $k$ performs the following operations:
\vspace{0.5mm}

\emph{1.\, Mobility update}:\, 
For $k \!\in\! \mathcal V_{{\rm M}}^n \rmv \cap \rmv \mathcal V_{{\rm M}}^{n-1}\!$, $\matr {\eta}^n_k$ and $\matr \Sigma^n_k$ are calculated from 
$(\matr {\mu}^{n-1}_k)^{[t^*]}\rmv$ and $(\matr C^{n-1}_k)^{[t^*]}\rmv$ as in \eqref{2d} (for MM2) or as in \eqref{2d_MM1} (for MM1). 
For $k\in\mathcal V_{{\rm M}}^n \rmv\setminus\rmv \mathcal V_{{\rm M}}^{n-1}\!$, $\matr {\eta}^n_k$ and $\matr \Sigma^n_k$ are the 
mean and covariance matrix of the Gaussian prior pdf $p(\matr x_{k}^n)$, which are assumed already available at agent $k$.

\vspace{.5mm}

\emph{2.\ Iterative message passing:} 
The message passing iterations are initialized ($t \rmv=\rmv 0$) with $(\matr {\mu}^{n}_k)^{[0]} \!\rmv=\rmv \matr {\eta}^n_k$ and 
$(\matr C^{n}_k)^{[0]} \!\rmv=\rmv \matr \Sigma^{n}_k$. At iteration $t \in \{1,\dots,t^*\}$, agent $k$ broadcasts $(\matr {\mu}_{\text{p},k}^{n})^{[t-1]}\rmv$ 
and $(c_{\text{p},k}^{n})^{[t-1]}\rmv$ and receives from the 
\pagebreak 
neighbors $(\matr {\mu}_{\text{p},l}^{n})^{[t-1]}\rmv$ and $(c_{\text{p},l}^{n})^{[t-1]}$, 
$l \!\in\! \cl{N}_k^{n}$. Note that the anchors 
($l \!\in\! \cl{N}_k^{n} \cap \cl{V}_{\rm{A}}^{n}$) broadcast their true position, so that $(\matr{\mu}^{n}_{\text{p},l})^{[t-1]} \rmv= \tilde{\matr{p}}_{l}^n$ and 
$(c^{n}_{\text{p},l})^{[t-1]}=0$. Then, new parameters $(\matr {\mu}^{n}_k)^{[t]}\rmv$ and $(\matr C^{n}_k)^{[t]}\rmv$ are calculated using the iterative 
belief approximation algorithm. After the last iteration ($t\rmv=\rmv t^*$), an approximation of the MMSE state estimate $\hat{\matr x}^{n}_{k}$ in \eqref{eq:mmse} is 
obtained as $(\matr {\mu}^{n}_k)^{[t^*]}$. This equals the result of \eqref{eq:mmse} with $p(\matr x_{k}^n\vert \matr d^{1:n})$ replaced by 
$\tilde{q}_k^{[t^*]}(\matr x_{k}^n)$.

The iterative belief approximation algorithm uses $\matr {\eta}^n_k$ and $\matr \Sigma^n_k$, which are locally available at agent $k$, 
and $(\matr {\mu}_{\text{p},l}^{n})^{[t-1]}\rmv$ and $(c_{\text{p},l}^{n})^{[t-1]}$, $l \rmv\in\rmv \cl{N}_k^{n}$, which were received from the neighbors 
of agent $k$. Therefore, at each message passing iteration $t$, each agent $k$ must broadcast to its neighbors $l \rmv\in\rmv \cl{N}_k^{n}$ 
only three real values, namely, two for $(\matr {\mu}_{\text{p},k}^{n})^{[t-1]}$ and one for $({c}_{\text{p},k}^{n})^{[t-1]}$.

\vspace{-1mm}

\section{Simulation Results}\label{sec:simRes}

We consider a region of interest (ROI) of size $120{\ist}\text{m} \times 120{\ist}\text{m}$ with the same $|\cl{V}_{\rm{M}}^{n}|\!=\!41$ agents 
and $|\cl{V}_{\rm{A}}^{n}|\!=\!18$ anchors at all $N \!=\! 30$ simulated time steps $n$. The anchors are regularly placed within the 
ROI. To avoid boundary effects, agents leaving the ROI reenter it at the respective opposite side. Agents and anchors have a communication radius of 
$20{\ist}\text{m}$; thereby, each agent communicates with one or two anchors. 
The agents measure distances according to \eqref{meas_kodel} with 
$\sigma_w \!=\! 1\text{m}$. 
For generating the agent trajectories, we set 
$T \!=\! 1\text{s}$, $\sigma_{v} \!=\!$\linebreak 
$\sqrt{1.5}{\ist}\text{m}/\text{s}$, and $\sigma_{a} \!=\!\sqrt{0.03}{\ist}\text{m}/\text{s}^2$. 
The initial agent positions are uniformly drawn on the ROI and, for MM2, the initial agent velocities 
are drawn from a Gaussian pdf with mean $[0 \,\, 0]^{\mathrm{T}}\rmv$ and covariance matrix $0.6 \cdot\rmv {\bf I}_2$.
For initializing the various algorithms, the prior pdf for $\matr p_{k}^{0}$ is chosen Gaussian with mean $\bm{\mu}^{0}_{\mathrm{p},k}$ and 
covariance matrix $900 \cdot\rmv {\bf I}_2$. Here, if agent $k$ is adjacent to one anchor $l$, then $\bm{\mu}^{0}_{\mathrm{p},k}$ is uniformly drawn 
from a circle of radius $d_{k,l}^{0}$ around the true anchor position $\tilde{\matr p}^{0}_{l}$, and if agent $k$ is adjacent
to two anchors $l$ and $l'\!$, then $\bm{\mu}^{0}_{\mathrm{p},k}$ is chosen as $(\tilde{\matr p}^{0}_{l} + \tilde{\matr p}^{0}_{l'})/2$. For MM2, the 
pdf for $\matr v_{k}^{0}$ is chosen Gaussian with mean $[ 0 \;\ist 0 ]^\text{T}\rmv$ and covariance matrix $0.6 \cdot\rmv {\bf I}_2$.

We compare the proposed hybrid BP--MF method as stated in Section \ref{sec:algorithmStatement} (abbreviated BPMF) with nonparametric 
BP (NBP) and sigma point BP (SBP). NBP \cite{lien} is an extension of the particle-based BP method of \cite{ihler} to mobile agents, and 
SBP \cite{meyer14sigma} is a low-complexity sigma-point-based BP scheme in which, similarly to BPMF, only Gaussian parameters are communicated. 
Our simulation of NBP uses 800 particles. For simulating BPMF, we perform the fixed-point iteration (with 30 iteration steps) multiple times 
\pagebreak 
with different initial values $\matr{\theta}_0$. More specifically, 20 initial values of $\matr{\mu}$
are drawn from ${m}_{k\to k}(\matr x_{k}^n)$, 20 are drawn from $\tilde{q}_k^{[t^*]}(\matr x_{k}^{n-1})$, and, for each adjacent anchor $l$, 20 are uniformly drawn 
from an annulus of radius $d_{k,l}^{n}$ and radial width $3 \ist \sigma_w$ around $\tilde{\matr p}^{n}_{l}$ \cite{ihler}. Furthermore, the initial values of
${c}_{\text{p}}$ and, for MM2, of ${c}_{\text{v}}$ and $c$ are always equal to the respective parameters of $\tilde{q}_k^{[t^*]}(\matr x_{k}^{n-1})$. 
Our measure of performance is the outage probability $P_{\text{out}} \triangleq \mathrm{Pr}\big[ \|\hat{\matr p}_{k}^{n} - \tilde{\matr p}_{k}^{n} \| \rmv>\rmv \tau\big]$, 
where $\tilde{\matr p}_{k}^{n}$ is the true position of agent $k$ at time $n$, $\hat{\matr p}_{k}^{n}$ is a corresponding estimate, and $\tau \rmv>\rmv 0$ is a threshold.

Fig.\ \ref{fig:results} shows the simulated outage probability $\hat{P}_{\text{out}}$, averaged over 30 simulation trials, of the three methods versus the 
outage threshold $\tau$. It is seen that, at $n \rmv=\! 1$, BPMF out\-performs NBP and SBP for $t^* \!=\rmv 30$; in particular, SBP performs poorly. Since 
BPMF and SBP use a Gaussian approximation, one may conclude that in the case of a noninformative prior (which is in force at $n \rmv=\! 1$), 
the Gaussian approximation degrades the performance of a pure BP scheme like SBP more than that of the proposed hybrid BP--MF scheme. 
At $n \rmv=\rmv 30$, for MM1, BPMF performs as NBP and SBP. However, for MM2, where the state can be predicted more accurately 
from the previous time, SBP outperforms both BPMF and NBP. Indeed, as previously observed in \cite{meyer14sigma}, SBP
works very well when informative prior knowledge is available. We expect that NBP would be similarly accurate if more particles were used; 
however, the complexity of SBP grows quadratically with the number of particles. It is also seen that for both MMs, contrary to BPMF, 
the performance of NBP and SBP at $n \rmv=\rmv 30$ does not improve when $t^*$ is increased beyond $5$. We note that in less dense networks, 
where beliefs can be multimodal, NBP can be expected to outperform SBP and BPMF.

The communication requirements, in terms of number of real values broadcast per message passing iteration $t$ by each agent $k$ to adjacent
agents $l \rmv\in\rmv \mathcal N_k^n$, are $3$ for BPMF, $5$ for SBP, and $1600$ for NBP.

\vspace{-2mm}

\section{Conclusion}

The proposed algorithm for cooperative localization and tracking combines the advantages of existing BP and MF methods: its accuracy is similar
to that of particle-based BP although only three real values per message passing iteration are broadcast by each agent, instead of hundreds of particles. 
Our simulations showed that the algorithm performs particularly well relative to pure BP-based methods when the prior information on the agent positions is imprecise.

\bibliographystyle{IEEEtran}
\bibliography{references}
\end{document}